\documentclass[aps,prd,floats,epsf,superscriptaddress]{revtex4}
\usepackage{graphicx}

\newcommand{\zpr}{\mbox{$Z'$}}

\def\bwt{\begin{widetext}}
\def\ewt{\end{widetext}}
\def\be{\begin{equation}}
\def\ee{\end{equation}}
\def\bea{\begin{eqnarray}}
\def\eea{\end{eqnarray}}
\def\bean{\begin{eqnarray*}}
\def\eean{\end{eqnarray*}}
\def\bary{\begin{array}}
\def\eary{\end{array}}

\def\bit{\begin{itemize}}
\def\eit{\end{itemize}}

\def\lan{\langle}
\def\ran{\rangle}

\def\lbar{\overline}

\def\nn{\nonumber}

\begin{document}

\rightline{\vbox{
    \hbox{MADPH-03-1350, UPR-1053T}
    \hbox{hep-ph/0310073}}}

\title{$Z'$ Mediated Flavor Changing Neutral Currents in $B$ Meson Decays}

\author{Vernon Barger}
\email[e-mail: ]{barger@oriole.physics.wisc.edu}
\affiliation{Department of Physics, University of Wisconsin, Madison, WI 53706}
\author{Cheng-Wei Chiang}
\email[e-mail: ]{chengwei@pheno.physics.wisc.edu}
\affiliation{Department of Physics, University of Wisconsin, Madison, WI 53706}
\author{Paul Langacker}
\email[e-mail: ]{pgl@electroweak.hep.upenn.edu}
\affiliation{Department of Physics and Astronomy,
University of Pennsylvania, Philadelphia, PA 19104}
\author{Hye-Sung Lee}
\email[e-mail: ]{comety@pheno.physics.wisc.edu}
\affiliation{Department of Physics, University of Wisconsin, Madison, WI 53706}

\date{\today}

\begin{abstract}
  We study the effects of an extra $U(1)'$ gauge boson with flavor changing
  couplings with fermion mass eigenstates on certain $B$ meson decays that are
  sensitive to such new physics contributions.  In particular, we examine to
  what extent the current data on $B_d \to \phi K$, $B_d \to \eta' K$ and $B_s
  \to \mu^+ \mu^-$ decays may be explained in such models, concentrating on the
  example in which the flavor changing couplings are left-chiral.  We find that
  within reasonable ranges of parameters, the $Z'$ contribution can readily
  account for the anomaly in $S_{\phi K_S}$ but is not sufficient to explain
  large branching ratio of $B_d \to \eta' K$ with the same parameter value.
  $S_{\phi K_S}$ and $S_{\eta' K_S}$ are seen to be the dominant observables
  that constrain the extra weak phase in the model.
\end{abstract}


\maketitle

%%%%%%%%%%%%%%%%%%%%%%%%%%%%%%%%%%%%%%%%%%%%%%%%%%%%%%%%%%%%%%%%%%%%%%%%%%%%%%%%
\section{Introduction \label{sec:intro}}
%%%%%%%%%%%%%%%%%%%%%%%%%%%%%%%%%%%%%%%%%%%%%%%%%%%%%%%%%%%%%%%%%%%%%%%%%%%%%%%%

$CP$ violation has been a puzzling phenomenon in the studies of elementary
particle physics since the first observation of its effects in hadronic kaon
decays almost four decades ago \cite{Christenson:fg}.  In the standard model
(SM), $CP$-violation is due entirely to the Cabibbo-Kobayashi-Maskawa (CKM)
mechanism \cite{Cabibbo:yz,Kobayashi:fv}, describing the mismatch between the
unitary transformations relating the up and down type quark mass eigenstates to
the corresponding weak eigenstates.  The CKM matrix involves a single weak
phase along with three mixing angles.  The validity of the CKM picture is
further strengthened by the fact that recent $\sin 2\beta$ measurements from
time-dependent $CP$ asymmetries of decay modes involving the $b \to c {\bar c}
s$ subprocess \cite{Browder:2003} agree well with the range of the weak phase
$\beta$ from many other constraints \cite{CKMFitter}. However, it is still
unknown whether there are any other sources that may give rise to
$CP$-violating effects.  Good places to search for deviations from the SM
predictions are decay processes that are expected to be rare in the SM, which
may reveal new physics through interference effects.  In particular,
discrepancies among the time-dependent $CP$ asymmetries of different $B$ decay
modes may show evidence for new physics
\cite{Grossman:1996ke,Fleischer:1996bv,London:1997zk,Fleischer:2001pc,%
Hiller:2002ci,Ciuchini:2002pd}.

Recently, an anomaly was reported in the time-dependent $CP$ asymmetry
measurement of the $B_d \to \phi K_S$ decay mode.  Within the framework of the
SM, this process should also provide us with information on the weak phase
$\beta$, up to about $5\%$ theoretical uncertainty
\cite{Grossman:1996ke,Grossman:1997gr}.  However, the averaged value of
$S_{\phi K_S}$ reported by the BaBar and Belle groups is \cite{Browder:2003}
(individual measurements to be quoted in Table \ref{tab:data})
\be
S_{\phi K_S} = -0.147 \pm 0.697 \; (S=2.11) ~.
\ee
This result is only about $1.3 \sigma$ away from the corresponding quantity
measured by the $B \to J/\psi K_S$ mode, $S_{J/\psi K_S} = 0.736 \pm 0.049$
\cite{Browder:2003}. However, the scale factor $S=2.11$ suggests a discrepancy
between the two experimental groups \footnote{see Ref.\ \cite{Hagiwara:fs} for
  the definition of the scale factor.}.  Before this discrepancy is settled,
the difference between $S_{\phi K_S}$ and $S_{J/\psi K_S}$ suggests the
possibility of new physics contributions.  From the theoretical point of view,
the $B \to \phi K_S$ decay is a loop-induced process involving $b \to s {\bar
  s} s$ penguin operators in the SM.  Therefore, it is susceptible to new
physics contributions even if they are suppressed by a large mass parameter
which characterizes the new physics scale.

In addition to model-independent approaches
\cite{Grossman:1996ke,Grossman:1997gr,Chiang:2003jn}, many studies have been
made to explain the anomaly in supersymmetric and related models
\cite{Datta:2002nr,Giri:2003jj}.  Such an effect can also be explained using
models in which the bottom quark is mixed with heavy mirror fermions with
masses of the order of the weak scale \cite{Morrissey:2003sc}.  It is the
purpose of this work to show that a new physics effect of similar size can be
obtained from some models with an extra $Z'$ boson.

$Z'$ bosons are known to naturally exist in well-motivated extensions of the SM
\cite{plrev}.  The $Z'$ mass is constrained by direct searches at Fermilab,
weak neutral current data, and precision studies at LEP and the SLC
\cite{Abe:1997fd,Erler:1999ub,afb}, which give a model-dependent lower bound
around 500 GeV.  The latter also severely limits the $Z$-$Z'$ mixing angle
$|\theta| < $ a few $\times 10^{-3}$.  A \zpr \ could be relevant to the NuTeV
experiment~\cite{NuTeV} and, if the couplings are not family
universal~\cite{Langacker:2000ju,Erler:1999ub}, to the anomalous value of the
forward-backward asymmetry $A^b_{FB}$~\cite{afb}.  (Earlier hints of a
discrepancy in atomic parity violation have largely disappeared due to improved
calculations of radiative corrections~\cite{Ginges:2003qt}.)  We therefore
study the $Z'$ boson in the mass range of a few hundred GeV to 1 TeV, assuming
no mixing between $Z$ and $Z'$.

Interesting phenomena arise when the $Z'$ couplings to physical fermion
eigenstates are non-diagonal.  This is possible if there exist additional
exotic fermions that have different $U(1)'$ charges from the ordinary fermions,
as found in $E_6$ models \cite{Nardi:1992nq,Babu:1996vt,Rizzo:1998ut}.
However, in these models left-handed fermion mixings may induce undesirable
flavor changing neutral currents (FCNC) mediated by the $Z$ boson even in the
absence of $Z$-$Z'$ mixing or nonuniversal family couplings.  One can avoid
this consequence by confining the mixing to be between right-handed fermions
and the exotic quarks \cite{Leroux:2001fx}.  Alternatively, other models give
family nonuniversal $Z'$ couplings as a result of different ways of constructing
families in some string models \cite{Chaudhuri:1994cd,cceepw,csu,csl}.  FCNC
and possibly new $CP$-violating phenomena will also occur in these models after
fermion mixings are taken into account. These can occur for both left and
right-handed fermions.

Although experiments on FCNC processes (such as the mass difference between
$K_L$ and $K_S$ and the $K_L \to \mu^+ \mu^-$ decay) have significantly
constrained the $Z'$ couplings of the first and second generation quarks to be
almost the same and diagonal, the couplings to the third generation are not
well constrained.  Similar statements apply to the charged leptons.  It has
been shown in Ref.\ \cite{Chaudhuri:1994cd,cceepw,csu,csl} that indeed the
third generation fermions can have different $Z'$ couplings from the other two
generations.

We use all of the above-mentioned features to study the imprints of the $Z'$
boson on certain processes that involve $b \to s$ transitions.  In Section
\ref{sec:formalism}, we present the model and framework to be studied.  In
Section \ref{sec:phiKs}, we show the constraints on the model parameters from
the current data of $S_{\phi K_S}$, $A_{\phi K_S}$ and the branching ratio
${\mathcal B}(B_d \to \phi K)$.  In Section \ref{sec:etaprKs}, we study a
related process $B_d \to \eta' K_S$, also including its $CP$ asymmetries and
branching ratio.  Another way to constrain the $b$-$s$-$Z'$ coupling is the
leptonic decay process $B_s \to \mu^+ \mu^-$.  We study this process in Section
\ref{sec:mumu} and show how sensitively Run II at Fermilab can probe this
coupling in the near future.  We conclude in Section \ref{sec:conclusions}.

%%%%%%%%%%%%%%%%%%%%%%%%%%%%%%%%%%%%%%%%%%%%%%%%%%%%%%%%%%%%%%%%%%%%%%%%%%%%%%%%
\section{Formalism \label{sec:formalism}}
%%%%%%%%%%%%%%%%%%%%%%%%%%%%%%%%%%%%%%%%%%%%%%%%%%%%%%%%%%%%%%%%%%%%%%%%%%%%%%%%

In this paper, we concentrate on models in which the interactions between the
$Z'$ boson and fermions are flavor nonuniversal for left-handed couplings and
flavor diagonal for right-handed couplings.  The analysis can be
straightforwardly extended to general cases in which the right-handed couplings
are also nonuniversal across generations.  The basic formalism of flavor
changing effects in the $Z'$ model with family nonuniversal and/or nondiagonal
couplings has been laid out in Ref.\ \cite{Langacker:2000ju}, to which we refer
readers for detail.  Here we just briefly review the ingredients needed in this
paper.

We write the $Z'$ term of the neutral-current Lagrangian in the gauge basis as
\be
\label{eq:Zpr}
{\cal L}^{Z'}
=-g' J'_{\mu} Z'^{\mu} ~,
\ee
where $g'$ is the gauge coupling associated with the $U(1)'$ group at the $M_W$
scale.  We neglect its renormalization group (RG) running between $M_W$ and
$M_{Z'}$.  The $Z'$ boson is assumed to have no mixing with the SM $Z$ boson
\footnote{Such mixings would modify the expressions for the phenomenological
  $\xi^{LL}$ and $\xi^{LR}$ parameters defined in Section
  \ref{sec:phiKs-amp-br}, but would not alter the discussion of the
  implications.}.  The chiral current is
\be
J'_{\mu}
= \sum_{i,j} {\lbar \psi_i^I} \gamma_{\mu}
  \left[ (\epsilon_{\psi_L})_{ij} P_L + (\epsilon_{\psi_R})_{ij} P_R \right]
  \psi^I_j ~,
\ee
where the sum extends over the flavors of fermion fields, the chirality
projection operators are $P_{L,R} \equiv (1 \mp \gamma_5) / 2$, the superscript
$I$ refers to the gauge interaction eigenstates, and $\epsilon_{\psi_L}$
($\epsilon_{\psi_R}$) denote the left-handed (right-handed) chiral couplings.
$\epsilon_{\psi_L}$ and $\epsilon_{\psi_R}$ are hermitian under the requirement
of a real Lagrangian.  The fermion Yukawa coupling matrices $Y_{\psi}$ in the
weak basis can be diagonalized as
\be
Y_{\psi}^D = V_{\psi_R} Y_{\psi} V_{\psi_L}^\dagger
\ee
using the bi-unitary matrices $V_{\psi_{L,R}}$ in $\psi_{L,R} = V_{\psi_{L,R}}
\psi_{L,R}^I$, where $\psi_{L,R}^I \equiv P_{L,R} \psi^I$ and $\psi_{L,R}$ are
the mass eigenstate fields.  The usual CKM matrix is then given by
\be
V_{\rm CKM} = V_{u_L} V_{d_L}^{\dagger} ~.
\ee
The chiral $Z'$ coupling matrices in the physical basis of down-type quarks
thus read
\bea
B^L_d
&\equiv& V_{d_L} \epsilon_{d_L} V_{d_L}^{\dagger} ~, \\
B^R_d
&\equiv& V_{d_R} \epsilon_{d_R} V_{d_R}^{\dagger} ~,
\eea
where the $B^{L,R}_d$ are hermitian.  We do not need the corresponding
couplings for up-type quarks or charged leptons in our discussions.

As long as $\epsilon_{d_{L,R}}$ is not proportional to the identity matrix,
$B^{L,R}_d$ will have nonzero off-diagonal elements that induce FCNC
interactions.  To see this, consider as an example the simplified
$\epsilon_{d_L}$ matrix for the down-type quarks of the form
\be
\epsilon_{d_L}
= Q_d
\left(\begin{array}{ccc}
1 & 0 & 0 \\
0 & 1 & 0 \\
0 & 0 & X
\end{array}\right) ~,
\label{eldiag}
\ee
where both $d$ and $s$ quarks have the same $Z'$ charge $Q_d$ and $X$ is the
ratio of the $Z'$ charge of $b$ to $Q_d$.  If we assume the mixing is among the
down-type quarks only, $V_{d_L}^{\dagger} = V_{\rm CKM}$ and $V_{u_L} = {\bf
  1}$.  Any redefinition of the quark fields by pure phase shifts would have no
effect on the resultant $B^L_d$.  All the off-diagonal couplings are
proportional to the $Z'$ charge difference between the $b$ quark and the $d,s$
quarks, as expected.  Using the standard parametrization \cite{Chau:fp}, the
explicit form of the off-diagonal $Z'$ coupling between $b$ and $s$ quarks, for
example, is
\be
B^L_{sb}
= (1-X) Q_d \cos\theta_{13} \cos\theta_{23}
  \left( \cos\theta_{12} \sin\theta_{23}
    + \cos\theta_{23} \sin\theta_{12} \sin\theta_{13} e^{-i \delta_{13}}
  \right) ~,
\ee
where $\theta_{ij}$ are the mixing angles between the $i$th and the $j$th
generations and $\delta_{13}$ is the $CP$-violating weak phase.  In this
example, $B^L_{sb}$ is proportional to the product $V_{tb} V_{ts}^*$ of
elements of the CKM matrix.

More generally, one can always pick a basis for the weak eigenstates in which
the $\epsilon_{d_{L,R}}$ matrices are diagonal and of the form (\ref{eldiag}),
though with different $Q_d$ and $X$ for the $\epsilon_{d_{L}}$ and
$\epsilon_{d_{R}}$. However, the Yukawa matrices $Y^u_\psi$ and $Y^d_\psi$ will
in general not be diagonal in that basis, so that $V_{u_L} \ne {\bf 1}$ and $V
_{d_L}^\dagger \ne V_{\rm CKM}$. In that case $B^{L,R}_d = V_{d_{L,R}}
\epsilon_{d_{L,R}} V_{d_{L,R}}^{\dagger} $ will in general be nondiagonal and
complex, with new mixing angles and $CP$ violating phases not directly related
to $V_{\rm CKM}$ \footnote{One could instead always work in the $V_{u_L} = {\bf
    1}$ basis, in which case $\epsilon_{d_{L,R}}$ would in general be
  off-diagonal and complex.}.

Instead of restricting ourselves to models with particular parameter choices in
the couplings and mixings, we will take the effective theory point of view and
constrain the effective couplings relevant to the decay modes of interest in
the following analysis.  However, to be more definite, we assume that the
right-handed coupling matrix $B^R_d$ is flavor diagonal.  If $B^R_d$ is
nondiagonal, new operators involving different chirality structures will be
induced in $B$ decays.

%%%%%%%%%%%%%%%%%%%%%%%%%%%%%%%%%%%%%%%%%%%%%%%%%%%%%%%%%%%%%%%%%%%%%%%%%%%%%%%%
\section{$B_d \to \phi K_S$ \label{sec:phiKs}}
%%%%%%%%%%%%%%%%%%%%%%%%%%%%%%%%%%%%%%%%%%%%%%%%%%%%%%%%%%%%%%%%%%%%%%%%%%%%%%%%

Within the SM, the $B^0 \to \phi K^0$ decay proceeds through the loop-induced
$b \to s {\bar s} s$ transition, which involves dominantly the QCD penguin but
also some electroweak (EW) and chromomagnetic penguin contributions.  To
illustrate possible modifications due to the existence of an extra $U(1)'$
gauge boson, we will neglect the smaller contributions from weak annihilation
diagram in the following analysis although they can play some role in enhancing
the branching ratios \cite{Cheng:2000hv}.  This two-body hadronic $B$ meson
decay can be conveniently analyzed in the framework of the effective weak
Hamiltonian and factorization formalism \cite{Buchalla:1995vs,Ali:1997nh}.

Since the penguin diagrams receive dominant contributions from the top quark
running in the loop, the effective Hamiltonian relevant for the charmless
$|\Delta S| = 1$ decays can be written as
\be
{\cal H}_{\rm eff}^{\rm SM}
= \frac{G_F}{\sqrt{2}}
  \left\{ V_{ub}V_{us}^* \left[ c_1 O_1 + c_2 O_2 \right]
  - V_{tb}V_{ts}^*
  \left[ \sum_{i=3}^{10} c_i O_i + c_g O_g \right] \right\}
  + \mbox{h.c.} ~.
\ee
Here
\be
O_1
= ({\bar u}_\alpha b_\alpha)_{V-A} ({\bar s}_\beta u_\beta)_{V-A} ~,
\qquad
O_2
= ({\bar u}_\alpha b_\beta)_{V-A} ({\bar s}_\beta u_\alpha)_{V-A}
\ee
are tree-level color-favored and color-suppressed operators,
\be
O_{3(5)}
= ({\bar s}_\alpha b_\alpha)_{V-A} ({\bar s}_\beta s_\beta)_{V-A (V+A)} ~,
\qquad
O_{4(6)}
= ({\bar s}_\alpha b_\beta)_{V-A} ({\bar s}_\beta s_\alpha)_{V-A (V+A)}
\ee
are the QCD penguin operators,
\be
O_{7(9)}
= \frac32 e_s
    ({\bar s}_\alpha b_\alpha)_{V-A} ({\bar s}_\beta s_\beta)_{V+A (V-A)} ~,
\qquad
O_{8(10)}
= \frac32 e_s
    ({\bar s}_\alpha b_\beta)_{V-A} ({\bar s}_\beta s_\alpha)_{V+A (V-A)}
\ee
are the EW penguin operators ($e_s = -1/3$ is the electric charge of the
strange quark ), and
\be
O_g
= \frac{g_s}{8 \pi^2} m_b
  {\bar s}_\alpha \sigma^{\mu\nu} T^a_{\alpha\beta} (1 + \gamma_5) b_\beta
    G_{\mu\nu}^a
\ee
is the chromomagnetic operator, where $({\bar q}_1 q_2)_{V \pm A} \equiv {\bar
  q}_1 \gamma_\mu (1 \pm \gamma_5) q_2$ and $\alpha,\beta$ refer to color
indices.

We mention in passing that the $Z'$ boson will also modify the $|\Delta B| = 2$
effective Hamiltonian relevant to $B_d$-$\bar B_d$ mixing, but in an
unnoticeable way.  This is because the additional contribution is proportional
to the square of the $Z'$ couplings between the first and third generations,
$|B^{L,R}_{db}|^2$, which is much more suppressed than the SM contribution.
Although the \zpr\ also contributes to $b \to (c \bar c) s$ transitions at the
tree level and gains a color factor relative to the SM tree process, it is
nevertheless suppressed by the $B^{L,R}_{sb}$ couplings and the \zpr\ mass in
comparison with $V_{cb}$ and the $W$ mass \cite{Atwood:2003tg}.  Consequently,
we do not study its effect on $\Delta M_{B_d}$ and $\sin 2\beta$ in charmed
modes.  Nevertheless, it can have significant effects on the $B_s$-$\bar B_s$
system if the couplings $B^{L,R}_{sb}$ are not too small, as we assume in the
current analysis.

%%%%%%%%%%%%%%%%%%%%%%%%%%%%%%%%%%%%%%%%%%%%%%%%%%%%%%%%%%%%%%%%%%%%%%%%%%%%%%%%
\subsection{Decay amplitude and branching ratio \label{sec:phiKs-amp-br}}
%%%%%%%%%%%%%%%%%%%%%%%%%%%%%%%%%%%%%%%%%%%%%%%%%%%%%%%%%%%%%%%%%%%%%%%%%%%%%%%%
  
In the generalized factorization approach \cite{Ali:1997nh}, the ${\bar B}_d
\to \phi {\bar K}^0$ decay amplitude is
\bea
\label{eq:sm-amp}
{\bar A}({\bar B}_d \to \phi {\bar K}^0)
&=& - \frac{G_F}{\sqrt{2}} V_{tb}V_{ts}^*
    \left[ a_3 + a_4 +a _5 -\frac12 \left( a_7 + a_9 + a_{10} \right) \right]
    X^{(BK,\phi)} ~,
\eea
where
\be
\label{eq:fact}
X^{(BK,\phi)}
\equiv \lan \phi | ({\bar s}_\alpha s_\alpha)_{V-A} | 0 \ran
       \lan {\bar K} | ({\bar s}_\beta b_\beta)_{V-A} | {\bar B} \ran
= 2 f_\phi m_\phi F_1^{BK}(m_\phi^2) (\epsilon^* \cdot p_B)
\ee
is a factorizable hadronic matrix element.  The coefficients $a_i$ are given by
\be
\label{eq:effwc}
a_{2i-1} = c^{\rm eff}_{2i-1} + \frac{1}{N_C^{\rm eff}} c^{\rm eff}_{2i} ~,
\qquad
a_{2i} = c^{\rm eff}_{2i} + \frac{1}{N_C^{\rm eff}} c^{\rm eff}_{2i-1} ~,
\ee
where $c^{\rm eff}_i$ are effective Wilson coefficients that should be used
when one replaces the one-loop hadronic matrix elements in the effective
Hamiltonian with the corresponding tree-level ones \cite{Ali:1997nh}.
Nonfactorizable effects are encoded in the effective number of colors $N_C^{\rm
  eff}$.  Throughout this paper, we take the naive choice $N_C^{\rm eff} = 3$
for illustration.

For the input parameters $\alpha_s(M_Z) = 0.118$, $\alpha_{\rm EM} = 1/128$;
the Wolfenstein parameters \cite{Wolfenstein:1983yz} $\lambda = 0.2240$, $A =
0.825$, $\rho = 0.21$ and $\eta = 0.34$ \cite{Battaglia:2003in};
$\sin^2\theta_W = 0.23$, $M_W = 80.42$ GeV; and the running quark masses $m_t =
168$ GeV, $m_b = 4.88$ GeV, $m_c = 1.5$ GeV, $m_s = 122$ MeV, $m_u = 4.2$ MeV,
and $m_d = 7.6$ MeV \cite{Xiao:2000qu}, the next-to-leading order (NLO)
effective Wilson coefficients \cite{Buchalla:1995vs,Ali:1997nh} for the
$|\Delta S| = 1$ weak Hamiltonian at the scale $\mu = 2.5$ GeV within the SM
are given in the second and third columns of Table \ref{tab:WC}.
%
%This is Table I
\begin{table*}
\caption{The SM Wilson coefficients used in the present analysis.  We
  assume the naive factorization for $a_i$ (i.e. $N_C^{\rm eff} = 3$), and
 ignore small differences between the $b \to s$ and ${\bar b} \to {\bar
  s}$ decays, expecting more significant effects from new physics.  $c^{\rm
  eff}_i$ and $a_i$ ($i = 3 \cdots 10$) should be multiplied by $10^{-5}$.
\label{tab:WC}}
\begin{ruledtabular}
\begin{tabular}{lll}
Operator & $c^{\rm eff}_i$ & $a_i (N_C^{\rm eff} = 3)$ \\
\hline
$O_1$     & $1.198$          & $1.064$ \\
$O_2$     & $-0.403$         & $-0.004$ \\
$O_3$     & $2817 + 301 i$   & $815$ \\
$O_4$     & $-6006 - 903 i$  & $-5067 - 803 i$ \\
$O_5$     & $2036 + 301 i$   & $-425$ \\
$O_6$     & $-7384 - 903 i$  & $-6705 - 803 i$ \\
$O_7$     & $-28 - 12 i$     & $-5 -12 i$ \\
$O_8$     & $70$             & $60 - 4 i$ \\
$O_9$     & $-1079 - 12 i$   & $-957 - 12 i$ \\
$O_{10}$  & $366$            & $6 - 4 i$
\end{tabular}
\end{ruledtabular}
\end{table*}

The $B_d \to \phi K^0$ decay width is given by
\be
\Gamma(B_d \to \phi K^0)
= \frac{p_c^3}{8\pi m_\phi^2} 
  \left| \frac{A(B_d \to \phi K^0)}{\epsilon^* \cdot p_B} \right|^2 ~,
\ee
where
\be
\label{eq:pc}
p_c
= \frac{\sqrt{\left[ m_B^2 - (m_\phi + m_K)^2 \right]
              \left[ m_B^2 - (m_\phi - m_K)^2 \right]}}{2m_B}
\ee
is the momentum of the decay particles in the center-of-mass frame.  With
$\tau_{B^0} = 1.534$ ps \cite{LEPBOSC}, $f_{\phi} = 237 $ MeV,
$F_1^{BK}(m_\phi^2) = 0.407$ \cite{Wirbel:1985ji} and meson masses given in
Ref.\ \cite{Hagiwara:fs}, the $CP$-averaged branching ratio in the SM is
\be
{\cal B}^{\rm SM}(B^0 \to \phi K^0)
\simeq 11 \times 10^{-6} ~.
\ee
This result is slightly above the $95\%$CL range of the current world average
value $(8.3 \pm 1.1) \times 10^{-6}$ given in Table \ref{tab:data}, but is
close to the previous calculation \cite{Giri:2003jj}.  (We ignore small
theoretical uncertainties in the SM here and in illustrating the consequences
of $Z'$ physics in the following sections.)

% This is Table II.
\begin{table*}
\caption{Experimental results of the $CP$-averaged branching ratios (quoted in
  units of $10^{-6}$) and $CP$ asymmetries of the $B \to \phi K$ and $B \to
  \eta' K$ decays.  References are given in square brackets.  The scale
  factor $S$ (defined in Ref.\ \cite{Hagiwara:fs}) is displayed in parentheses
  when it is larger than 1.
\label{tab:data}}
\begin{ruledtabular}
%\footnotesize
\begin{tabular}{lllll}
 Mode & BaBar & Belle & CLEO & Avg. \\
\hline
${\mathcal B}(B^0 \to \phi K^0)$
        & $8.4^{+1.5}_{-1.3} \pm 0.5$ \cite{Aubert:2003hz}
        & $9.0^{+2.2}_{-1.8} \pm 0.7$ \cite{Chen:2003jf}
        & $5.4^{+3.7}_{-2.7} \pm 0.7$ \cite{Briere:2001ue} %(<12.3)
        & $8.3 \pm 1.1$ \\
$S_{\phi K_S}$
        & $0.45\pm0.43\pm0.07$ \cite{Browder:2003}
        & $-0.96\pm0.50^{+0.09}_{-0.11}$ \cite{Browder:2003}
        & -
        & $-0.147\pm0.697 \, (S=2.11)$ \\
$A_{\phi K_S}$
        & $0.38\pm0.37\pm0.12$ \cite{Browder:2003}
        & $-0.15\pm0.29\pm0.07$ \cite{Browder:2003}
        & -
        & $0.046\pm0.256 \, (S=1.08)$ \\
\hline
${\mathcal B}(B^0 \to \eta' K^0)$
        & $60.6\pm5.6\pm4.6$ \cite{Aubert:2003bq}
        & $68\pm10^{+9}_{-8}$ \cite{Aihara} 
        & $89^{+18}_{-16}\pm9$ \cite{Richichi:1999kj}
        & $65.18\pm6.18 \, (S=1.03)$ \\
$S_{\eta' K_S}$ 
        & $0.02\pm0.34\pm0.03$ \cite{Browder:2003}
        & $0.43\pm0.27\pm0.05$ \cite{Browder:2003}
        & -
        & $0.269\pm0.214$ \\
$A_{\eta' K_S}$
        & $-0.10\pm0.22\pm0.03$ \cite{Browder:2003}
        & $-0.01\pm0.16\pm0.04$ \cite{Browder:2003}
        & -
        & $-0.042\pm0.132$ \\
\end{tabular}
\end{ruledtabular}
\end{table*}

With FCNC, the $Z'$ boson contributes at tree level, and its contribution will
interfere with the standard model contributions.  In particular, the
flavor-changing couplings of the $Z'$ with the left-handed fermions will
contribute to the $O_9$ and $O_7$ operators for left (right)-handed couplings
at the flavor-conserving vertex, i.e., $c_{9,7}(M_W)$ receive new contributions
from $Z'$.  On the other hand, the right-handed flavor changing couplings yield
new operators with coefficients that contain another weak phase associated with
$B^R_{sb}$.  We will ignore these contributions in this paper.

The effective Hamiltonian of the $b \to s {\bar s} s$ transition mediated by
the $Z'$ is
\be
{\cal H}_{\rm eff}^{Z'}
= - \frac{4 G_F}{\sqrt{2}}
    \left(\frac{g' M_Z}{g_Y M_{Z'}}\right)^2 B^L_{sb}
    \left( B^L_{ss} O_9 + B^R_{ss} O_{7} \right)
  + \mbox{h.c.} ~,
\ee
where $g_Y = e / (\sin \theta_W \cos \theta_W)$, and $B^L_{ij}$ and $B^R_{ij}$
refer to the left- and right-handed effective $Z'$ couplings of the quarks $i$
and $j$ at the weak scale, respectively.  The diagonal elements are real due to
the hermiticity of the effective Hamiltonian, but the off-diagonal elements may
contain weak phases.  Only one new weak phase associated with $B^L_{sb}$ can be
introduced into the theory under our assumption of neglecting $B^R_{sb}$. We
denote this by $\phi_L$ and write $B^L_{sb} = |B^L_{sb}| e^{i \phi_L}$.  As
${\cal H}_{\rm eff}^{Z'}$ has the same operators $O_9$ and $O_{7}$ as in the SM
effective Hamiltonian, the strong phases from long-distance physics should be
the same.

Since heavy degrees of freedom in the theory have already been integrated out
at the scale of $M_W$, the RG evolution of the Wilson coefficients after
including the new contributions from $Z'$ is exactly the same as in the SM.  We
obtain the branching ratio
\be
\label{eq:BRphiK0}
{\cal B}^{{\rm SM+}Z'}({\bar B} \to \phi {\bar K}^0)
\simeq {\cal B}^{\rm SM}({\bar B} \to \phi {\bar K}^0)
       \left| 1 - 
         \left[ (41.8 - 7.1 i) \xi^{LL} + (46.2 - 8.6 i) \xi^{LR} \right]
         e^{i \phi_L}
       \right|^2 ~,
\ee
where
\bea
\xi^{LL}
&\equiv& \left| \left(\frac{g' M_Z}{g_Y M_{Z'}}\right)^2
         \frac{B^L_{sb} B^L_{ss}}{V_{tb} V_{ts}^*} \right| ~, \\
\xi^{LR}
&\equiv& \left| \left(\frac{g' M_Z}{g_Y M_{Z'}}\right)^2
         \frac{B^L_{sb} B^R_{ss}}{V_{tb} V_{ts}^*} \right| ~,
\eea
and $|V_{tb} V_{ts}^*| \simeq 0.04$.  The second and third terms in
Eq.~(\ref{eq:BRphiK0}) represent the $Z'$ contributions from left- and
right-handed couplings with the $s {\bar s}$ in the final state, respectively.
We have assumed for definiteness that $B^L_{ss}$ and $B^R_{ss}$ have the same
sign, so that the $\xi^{LL}$ and $\xi^{LR}$ terms interfere constructively.
The branching ratio predicted by our model depends on the absolute ratios
$\xi^{LL}$ and $\xi^{LR}$ and the weak phase $\phi_L$.

\begin{figure}[h]
\includegraphics[width=.4\textwidth]{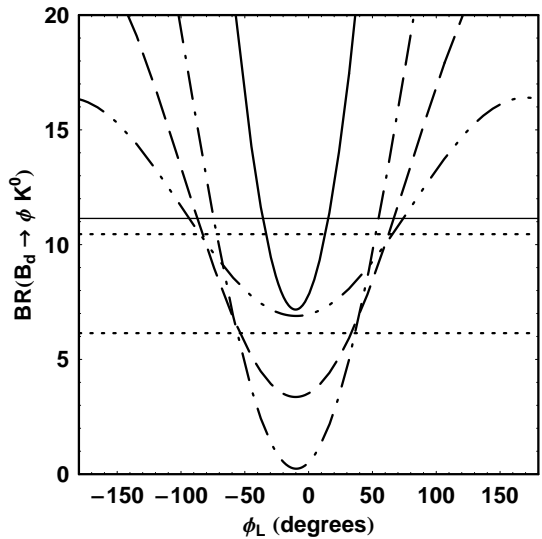}
\caption{The branching ratio ${\cal B}^{{\rm SM+}Z'}(B \to \phi K^0)$ in units
  of $10^{-6}$ versus the weak phase $\phi_L$ associated with the effective
  $Z'$ coupling $B^L_{sb}$.  The current experimental range at $95\%$CL is
  shown by the two horizontal dotted lines.  The SM prediction is the thin
  horizontal line.  The thick solid and dashed curves include both left-handed
  and right-handed couplings with $\xi^{LL} = \xi^{LR} = 0.02$ and $0.005$,
  respectively.  The single-dot-dashed and double-dot-dashed curves involve
  only the left-handed couplings with $\xi^{LL} = 0.02$ and $0.005$,
  respectively.
\label{fig:BRphiKs}}
\end{figure}

We show the branching ratios as a function of $\phi_L$ in
Fig.~\ref{fig:BRphiKs}.  Generically, one expects a ratio $g'/g_Y \sim {\cal
  O}(1)$ and $M_{Z'}$ to be a few to around 10 times $M_Z$.  We assume that the
product $|B^L_{sb} B^L_{ss}|$ is numerically about the same as $|V_{tb}
V_{ts}^*|$, and take $\xi^{LL} = 0.02$ and $0.005$ as representative values for
numerical analyses in this and the following sections \footnote{One may want to
  compare our choices of $\xi^{LL}$ here with the bounds on the product of the
  effective $s$-$b$-$Z'$ coupling and the ratio $M_Z^2 / M_{Z'}^2$ obtained
  from certain semileptonic $B$ decays in Ref.\ \cite{Leroux:2001fx}, although
  a specific $E_6$ model with leptophobic features is assumed in their study.}.
It is straightforward to scale the results to other $\xi^{LL}$ values.

% This is Figure 1

To quantify the effects of right-handed couplings, we consider $\xi^{LR} =
0.02$ and $0.005$ and show the corresponding curves in Fig.~\ref{fig:BRphiKs}.
The branching ratio curves are almost symmetric about $\phi_L = 0$, with the
slight asymmetry set by the small strong phases in the Wilson coefficients.
This echos the fact that the contributing amplitudes in Eq.~(\ref{eq:BRphiK0})
have the largest constructive interference when $\phi_L \simeq 0$.  To be
consistent with the measured branching ratio of $B^0 \to \phi K^0$, our weak
phase $\phi_L$ in the region $-80^{\circ} \sim 60^{\circ}$ is favored, with the
exact range depending upon $\xi^{LL}$ and $\xi^{LR}$ in the model.  For some
parameter choices, it can leave us a two-fold ambiguity, which can be resolved
using further information to be discussed in the following subsection.

%%%%%%%%%%%%%%%%%%%%%%%%%%%%%%%%%%%%%%%%%%%%%%%%%%%%%%%%%%%%%%%%%%%%%%%%%%%%%%%%
\subsection{Time-dependent $CP$ asymmetries \label{sec:phiKs-CPA}}
%%%%%%%%%%%%%%%%%%%%%%%%%%%%%%%%%%%%%%%%%%%%%%%%%%%%%%%%%%%%%%%%%%%%%%%%%%%%%%%%

The time-dependent $CP$ asymmetry for $B \to \phi K_S$ is
\bea
a_{\phi K_S}(t)
&=& \frac{\Gamma(\lbar{B}^0(t)\to\phi K_S)-\Gamma(B^0(t)\to\phi K_S)}
         {\Gamma(\lbar{B}^0(t)\to\phi K_S)+\Gamma(B^0(t)\to\phi K_S)}
\nn \\
&=& A_{\phi K_S} \cos (\Delta M_{B_d}t) + S_{\phi K_S} \sin (\Delta M_{B_d}t),
\eea
where the direct and the indirect $CP$ asymmetry parameters are given
respectively by
\be
A_{\phi K_S}
= \frac{|\lambda_{\phi K_S}|^2-1}{|\lambda_{\phi K_S}|^2+1} ~,
\qquad
S_{\phi K_S}
= \frac{2Im \left[ \lambda_{\phi K_S} \right]}{|\lambda_{\phi K_S}|^2+1}.
\ee
The parameter $\lambda_{\phi K_S}$ is defined by
\be
\label{eq:lambda_phiKs}
\lambda_{\phi K_S}
\equiv \eta_{\phi K_S} 
       \left( \frac{q}{p} \right)_B \left( \frac{p}{q} \right)_K
       \frac{\lbar{A}(\phi \lbar{K}^0)}{A(\phi K^0)} ~,
\ee
where $\eta_{\phi K_S} = -1$ is the $CP$ eigenvalue of the $\phi K_S$ state,
and
\be
\left( \frac{q}{p} \right)_B = \frac{V_{tb}^* V_{td}}{V_{tb} V_{td}^*} ~,
\qquad
\left( \frac{p}{q} \right)_K = \frac{V_{cs} V_{cd}^*}{V_{cs}^* V_{cd}}
\ee
are factors that account for the mixing effects in neutral $B$ and $K$ meson
systems, respectively.

% This is Figure 2
\begin{figure}[h]
\centerline{(a) \hspace{7.5cm} (b)}
\includegraphics[width=.4\textwidth]{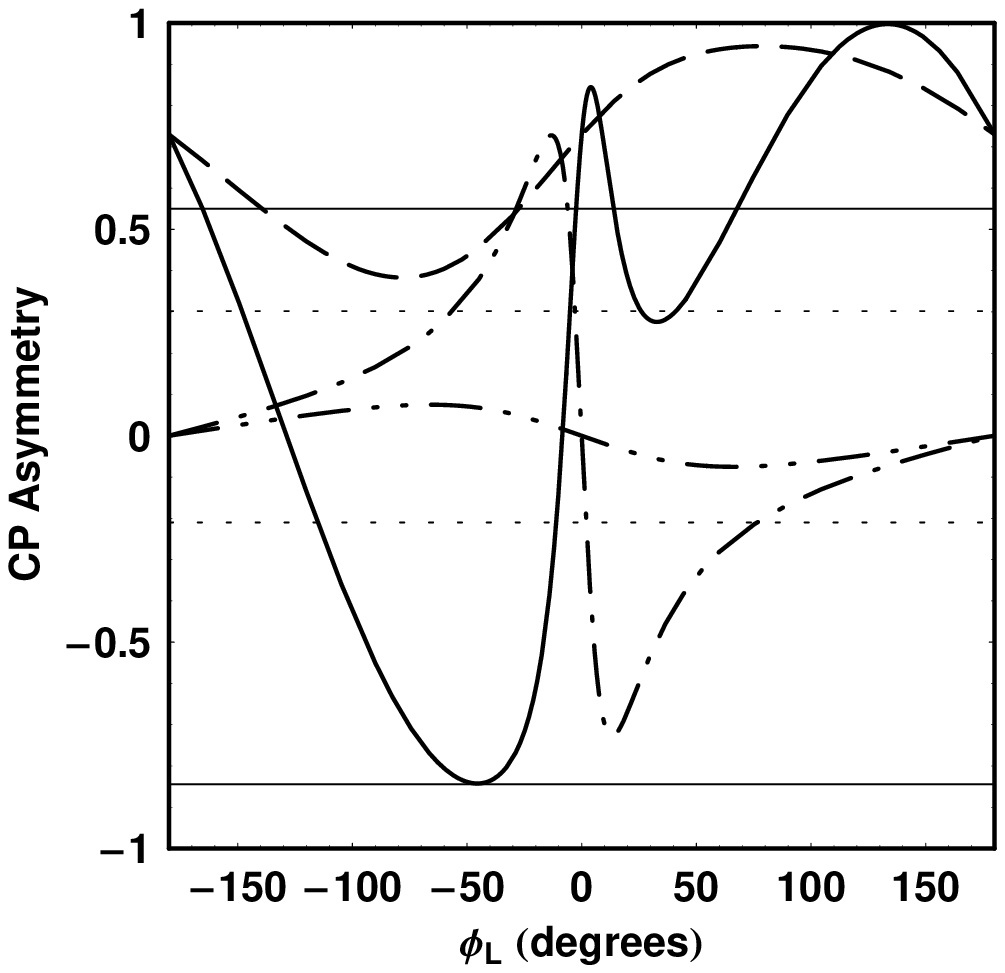}
\hspace{1cm}
\includegraphics[width=.4\textwidth]{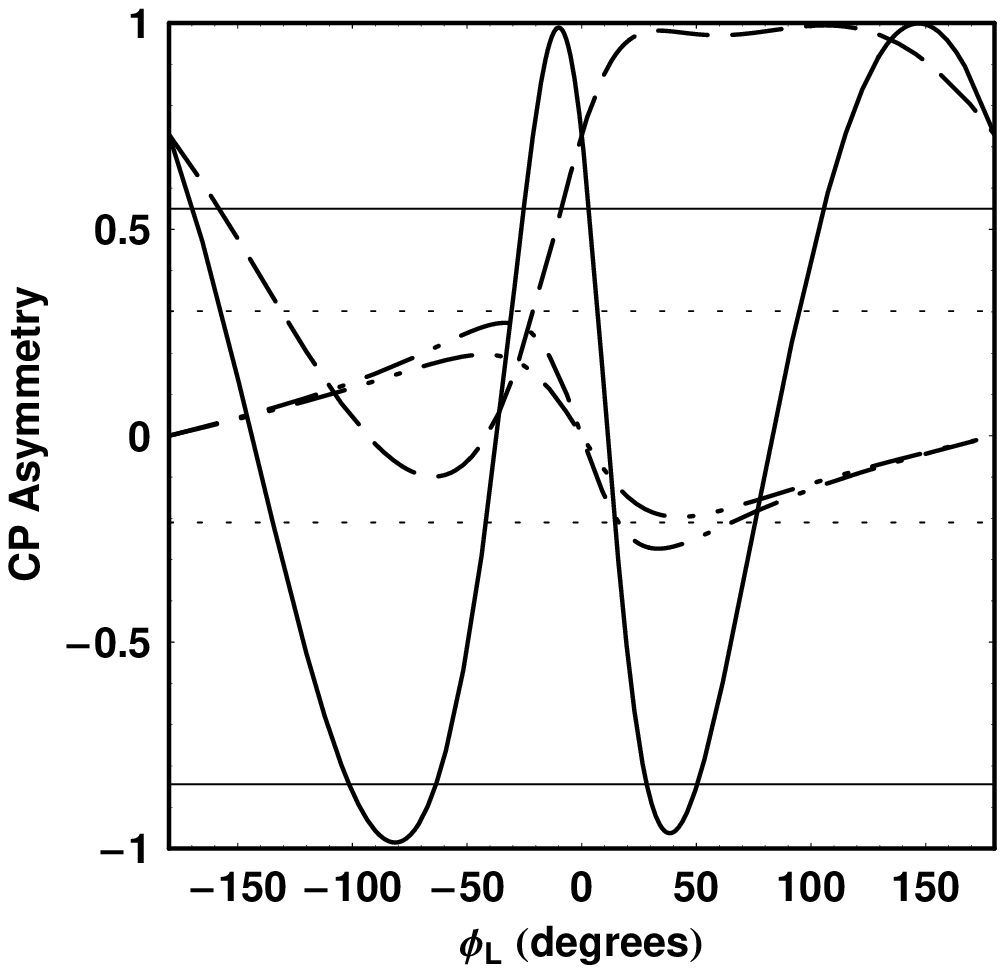}
\caption{The time-dependent $CP$ asymmetries, $S_{\phi K_S}$ and $A_{\phi
    K_S}$, versus $\phi_L$.  The current experimental ranges at $1\sigma$ level
  are shown by the thin horizontal solid and dotted lines, respectively.  The
  SM predicts $(S_{\phi K_S},A_{\phi K_S}) \simeq (0.73,0)$ (not shown).  The
  thick solid and dashed curves are $S_{\phi K_S}$ for $\xi^{LL} = 0.02$ and
  $0.005$, respectively.  The corresponding curves for $A_{\phi K_S}$ are
  displayed using single-dotted and double-dotted dashes.  Plot (a) has 
  $\xi^{LR} = 0$; plot (b) has $\xi^{LR} = \xi^{LL} = 0.02$
  and $0.005$.
\label{fig:ACPphiKs}}
\end{figure}

We show our estimates of $S_{\phi K_S}$ and $A_{\phi K_S}$ as a function of the
new weak phase $\phi_L$ in Fig.~\ref{fig:ACPphiKs}.  In
Fig.~\ref{fig:ACPphiKs}(a), we have choices $\xi^{LL} = 0.02$ and $0.005$, but
set $\xi^{LR} = 0$.  The SM prediction of $S_{\phi K_S}$ and $A_{\phi K_S}$ are
$0.73$ and $0$, respectively.  We see that the measured $A_{\phi K_S}$ does not
give much constraint on the weak phase $\phi_L$, except for the regions between
$-55^{\circ} \sim 80^{\circ}$ when $\xi^{LL} = 0.02$.  The $S_{\phi K_S}$ data
can be readily fitted within $1\sigma$ for values of $\xi^{LL}$ chosen here.

In Fig.~\ref{fig:ACPphiKs}(b), we turn on the right-handed couplings and set
$\xi^{LR} = \xi^{LL}$.  We notice that the variation of $A_{\phi K_S}$ is
within the experimental $1\sigma$ limits for the most range of $\phi_L$.  As
illustrated by the thick solid curve in Fig.~\ref{fig:ACPphiKs}(b), there are
four possible ranges of $\phi_L$ that can fit the averaged $S_{\phi K_S}$ if
both $\xi^{LL}$ and $\xi^{LR}$ are large enough.  For smaller $\xi^{LL}$ and
$\xi^{LR}$, however, only a region of negative $\phi_L$ is favored.

In order to satisfy both CP asymmetry constraints, $\phi_L$ should have
negative value in most cases.  Only $\xi^{LL} = \xi^{LR} = 0.02$ can have some
positive $\phi_L$ range.  Combining the constraints from ${\cal B}(B \to \phi
K)$ at $95\%$CL and both $S_{\phi K_S}$ and $A_{\phi K_S}$ at $1\sigma$ level,
we find the following allowed regions of $\phi_L$.  If we take $\xi^{LR} =
\xi^{LL} = 0.02$, $5^{\circ} \alt \phi_L \alt 15^{\circ}$ is favored.  If we
take $\xi^{LR} = \xi^{LL} = 0.005$, $-80^{\circ} \alt \phi_L \alt -55^{\circ}$
is favored.  If we take $\xi^{LL} = 0.02$ and ignore $\xi^{LR}$, then
$-70^{\circ} \alt \phi_L \alt -55^{\circ}$ is favored.  For $\xi^{LL} = 0.005$
and $\xi^{LR} = 0$, $-80^{\circ} \alt \phi_L \alt -30^{\circ}$ is favored.

%%%%%%%%%%%%%%%%%%%%%%%%%%%%%%%%%%%%%%%%%%%%%%%%%%%%%%%%%%%%%%%%%%%%%%%%%%%%%%%%
\section{$B_d \to \eta' K_S$ \label{sec:etaprKs}}
%%%%%%%%%%%%%%%%%%%%%%%%%%%%%%%%%%%%%%%%%%%%%%%%%%%%%%%%%%%%%%%%%%%%%%%%%%%%%%%%

The $B_d \to \eta' K_S$ is another decay mode whose time-dependent $CP$
asymmetry $S_{\eta' K_S}$ is expected to give us the SM $\sin 2\beta$.  Current
data reported by BaBar and Belle (see Table \ref{tab:data}) are both lower than
the SM prediction, although consistent within $2\sigma$.  Since this process
also contains the $O_7$ and $O_9$ operators in the amplitude, we discuss the
$Z'$ effects on its observables.

The perturbative calculations of the $B \to \eta' K$ branching ratios are
significantly smaller than the observed values.  This discrepancy can be
explained by adding a singlet-penguin amplitude, where $\eta'$ is produced
through a flavor-singlet neutral current, to interfere constructively with the
QCD penguin contributions \cite{Lipkin:1990us,Chiang:2001ir}.  Another analysis
\cite{Beneke:2002jn} found that it is hard to obtain a sizeable flavor-singlet
amplitude from perturbative calculations, but QCD penguin amplitudes can be
enhanced by an asymmetric treatment of the $s{\bar s}$ component of the $\eta'$
wavefunction.  Since this matter is still debatable, we will follow the usual
effective Hamiltonian approach \cite{Buchalla:1995vs,Ali:1997nh} and put the
emphasis on what kind of effects the $Z'$ boson may provide.

Following the notation in Ref.\ \cite{Giri:2003jj}, the decay amplitude of
${\bar B}^0 \to \eta' {\bar K}^0$ can be written as
\bea
\label{eq:etaprK-sm-amp}
&& {\bar A}({\bar B}^0 \to \eta' {\bar K}^0) \nn \\
&& = i \frac{G_F}{\sqrt{2}}
    \left[ V_{ub} V_{us}^* a_2 X_2 - V_{tb} V_{ts}^*
      \left\{ \left[ a_4 - \frac{a_{10}}{2}
             + \left( a_6 - \frac{a_8}{2} \right) R_1 \right] X_1
             + \left[ 2(a_3 - a_5) - \frac12 (a_7 - a_9) \right] X_2
         \right.\right. \nn \\
&& \qquad\quad
      \left.\left.
             + \left[ a_3 + a_4 - a_5 + \frac12 (a_7 - a_9 - a_{10})
                 + \left( a_6 - \frac{a_8}{2} \right) R_2 \right] X_3
      \right\} \right] ~,
\eea
where
\bea
X_1
&=& i(m_B^2 - m_{\eta'}^2) \frac{X_{\eta'}}{\sqrt{2}}
    f_K F_0^{B\pi}(m_{K^0}^2) ~, \nn \\
X_2
&=& i(m_B^2 - m_{K^0}^2) \frac{X_{\eta'}}{\sqrt{2}}
    f_{\pi} F_0^{BK}(m_{\eta'}^2) ~, \nn \\
X_3
&=& i(m_B^2 - m_{K^0}^2) Y_{\eta'}
    \sqrt{2f_K^2 - f_{\pi}^2} F_0^{BK}(m_{\eta'}^2) ~, \nn \\
R_1
&=& \frac{2 m_{K^0}^2}{(m_b - m_d)(m_s + m_d)} ~, \nn \\
R_2
&=& \frac{2(2 m_{K^0}^2 - m_{\pi}^2)}{2 m_s (m_b - m_s)} ~,
\eea
and $X_{\eta'} = 0.57$ and $Y_{\eta'} = 0.82$ are mixing parameters for the
choice of the $\eta'$ meson wavefunction to be $(2 s{\bar s} + u{\bar u} +
d{\bar d})/\sqrt{6}$.

Since the $B \to \eta' K$ has two pseudoscalar mesons in the final state, the
decay width is
\be
\Gamma(B_d \to \eta' K^0)
= \frac{p_c}{8\pi m_B^2} \left| A(B_d \to \eta' K^0) \right|^2 ~,
\ee
where $p_c$ is defined in a similar way to Eq.~(\ref{eq:pc}).  With $f_{\pi} =
131$ MeV and $f_K = 159.8$ MeV \cite{Hagiwara:fs}, $F_0^{B\pi}(m_{K^0}^2) =
0.335$ and $F_0^{BK}(m_{\eta'}^2) = 0.391$ \cite{Wirbel:1985ji}, we have ${\cal
  B}^{\rm SM}(B \to \eta' K^0) \simeq 38 \times 10^{-6}$, which is much lower
than the experimental average of $(65.18\pm6.18) \times 10^{-6}$ (see Table
\ref{tab:data}).

As in the case of $B \to \phi K$ decays, our model makes extra contributions to
$O_9$ and $O_{7}$ at the weak scale.  The branching ratio is
\be
{\cal B}^{{\rm SM}+Z'}({\bar B} \to \eta' {\bar K}^0)
\simeq {\cal B}^{\rm SM}({\bar B} \to \eta' {\bar K}^0)
       \left| 1 -
         \left[ (7.0 - 0.5 i) \xi^{LL} + (2.9 - 0.4 i) \xi^{LR} \right]
         e^{i \phi_L}
       \right|^2 ~.
\ee
We notice that the coefficient of $\xi^{LL}$ and that of $\xi^{LR}$ also tend
to have constructive interference between themselves according to our
assumption that $B^L_{ss}$ and $B^R_{ss}$ have the same sign.  The magnitudes
of these coefficients, however, are much smaller than those in
Eq.~(\ref{eq:BRphiK0}).  This is simply because the terms that receive
contributions from the $Z'$ boson (mostly $a_9$) have some cancellation between
the $X_2$ and $X_3$ terms in Eq.~(\ref{eq:etaprK-sm-amp}).  These observations
qualitatively tell us why the $\eta' K$ decays are not affected quite as much
by the $Z'$ effects.

% This is Figure 3
\begin{figure}[h]
\includegraphics[width=.4\textwidth]{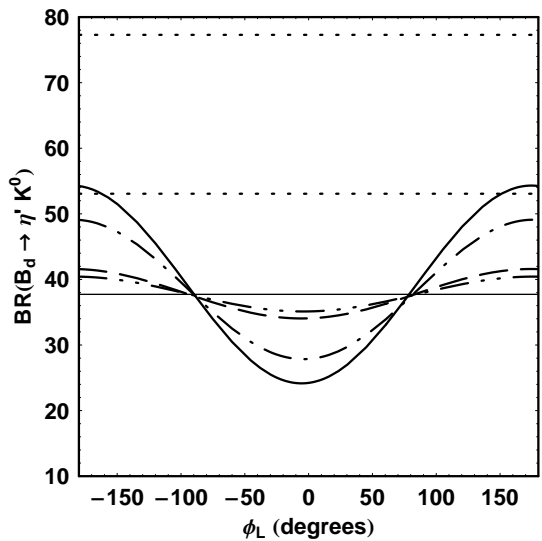}
\caption{The branching ratio ${\cal B}^{{\rm SM}+Z'}(B \to \eta' K^0)$ in units
  of $10^{-6}$ versus $\phi_L$.  The current experimental range at $95\%$CL is
  shown by the two horizontal dotted lines.  The SM prediction is the thin
  horizontal line.  The thick solid and dashed curves include both left-handed
  and right-handed couplings with $\xi^{LL} = \xi^{LR} = 0.02$ and $0.005$,
  respectively.  The single-dot-dashed and double-dot-dashed curves involve
  only the left-handed couplings with $\xi^{LL} = 0.02$ and $0.005$,
  respectively.
\label{fig:BRetaprKs}}
\end{figure}

We see in Fig.~\ref{fig:BRetaprKs} that the $Z'$ boson can explain the gap
between the observed branching ratio and the SM prediction only around $\phi_L
= \pm 180^{\circ}$ even with large couplings in both $\xi^{LL}$ and $\xi^{LR}$.
As we will see, however, this region is not favored by the CP asymmetry
constraints.  Therefore, we must attribute this anomaly to some other unknown
source.

% This is Figure 4
\begin{figure}[h]
\centerline{(a) \hspace{7.5cm} (b)}
\includegraphics[width=.4\textwidth]{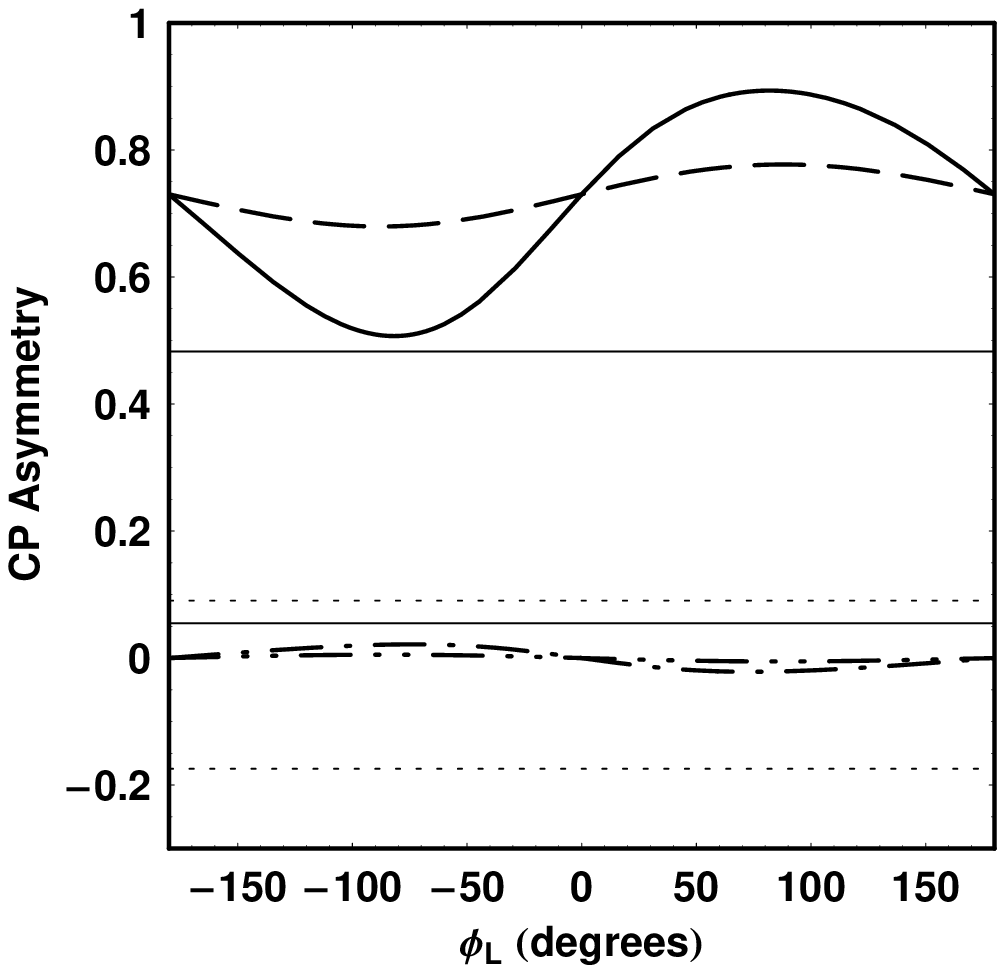}
\hspace{1cm}
\includegraphics[width=.4\textwidth]{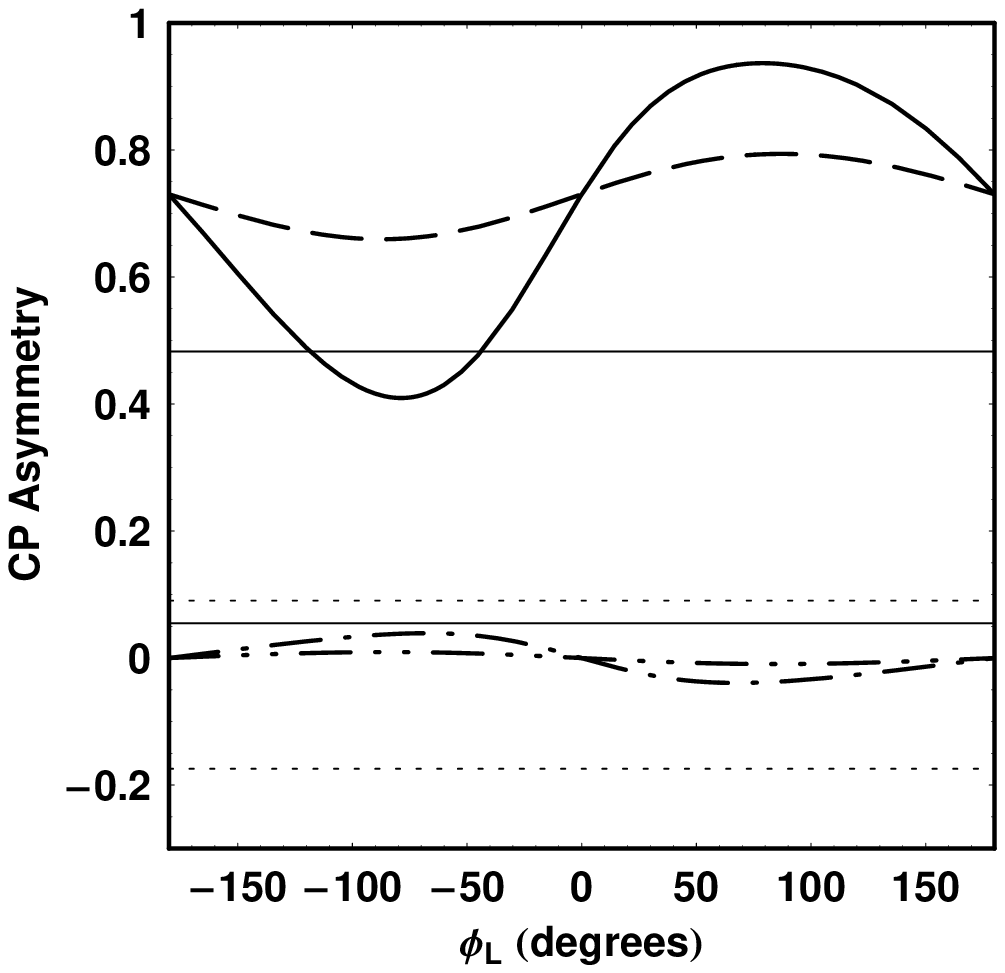}
\caption{The time-dependent $CP$ asymmetries, $S_{\eta' K_S}$ and $A_{\eta'
    K_S}$, versus $\phi_L$.  The current experimental ranges at $1\sigma$ level
  are shown by the thin horizontal solid and dotted lines, respectively.  The
  SM predicts $(S_{\eta' K_S},A_{\eta' K_S}) \simeq (0.73,0)$ (not shown).  The
  thick solid and dash-dotted curves are $S_{\eta' K_S}$ for $\xi^{LL} = 0.02$
  and $0.005$, respectively.  The corresponding curves for $A_{\eta' K_S}$ are
  displayed using single-dotted and double-dotted dashes.  Plot (a) has 
  $\xi^{LR} = 0$; plot (b) has $\xi^{LR} = \xi^{LL} = 0.02$
  and $0.005$.
\label{fig:ACPetaprKs}}
\end{figure}

The asymmetry curves for $B_d \to \eta' K_S$ are shown in
Fig.~\ref{fig:ACPetaprKs}.  We do not get useful constraints from current data
on $A_{\eta' K_S}$.  The value of $A_{\eta' K_S}$ does not vary much from its
SM prediction throughout the whole range of $\phi_L$.  The averaged value of
$S_{\eta' K_S}$ can be explained at $1\sigma$ level by simultaneously taking
large values of both left- and right-handed couplings (the solid curve in
Fig.~\ref{fig:ACPetaprKs}(b)).  In this case, however, only negative $\phi_L$
around $-120^{\circ} \sim -40^{\circ}$ is favored from the $S_{\eta' K_S}$
constraint. Other cases do not explain the $S_{\eta' K_S}$ anomaly though all
of them favor negative value of $\phi_L$ to approach the $1 \sigma$ limit.

Leaving $95\%$CL of branching ratio constraints, we have only $\xi^{LL} =
\xi^{LR} = 0.02$ case that can satisfy both $B \to \eta' K$ and $B \to \eta'
K_S$ CP asymmetry with a two-fold range of $\phi_L$, $-120^{\circ} \sim
-100^{\circ}$ and $-60^{\circ} \sim -40^{\circ}$. Attributing the branching
ratio of $B \to \eta' K$ to some unknown effects, the latter is favored by the
$B \to \phi K$ branching ratio.

%%%%%%%%%%%%%%%%%%%%%%%%%%%%%%%%%%%%%%%%%%%%%%%%%%%%%%%%%%%%%%%%%%%%%%%%%%%%%%%%
\section{$B_s \to \mu^+ \mu^-$ \label{sec:mumu}}
%%%%%%%%%%%%%%%%%%%%%%%%%%%%%%%%%%%%%%%%%%%%%%%%%%%%%%%%%%%%%%%%%%%%%%%%%%%%%%%%

$B \to \ell^+ \ell^-$ decays are good candidates to observe FCNC interactions
beyond the SM.  In the SM, the branching ratio of such processes are
proportional to $m_{\ell}^2$ and the corresponding CKM factors.  Therefore,
$B_s$ decays in general have branching ratios larger than the corresponding
$B_d$ decays by a factor of $\sim \lambda^2$.  Although $B_s \to \tau^+ \tau^-$
should have a still larger branching ratio, this mode is hard to detect at
hadron colliders.  The best mode for experimental study is thus $B_s \to \mu^+
\mu^-$.

Recently, studies in some minimal supersymmetric standard model (MSSM) and
minimal supergravity (mSUGRA) models predict that the branching ratio of the
rare decay $B_s \to \mu^+ \mu^-$ can be large enough to be observable by
Tevatron Run II \cite{Babu:1999hn}.  Using 113 pb$^{-1}$ of data from Run II,
the CDF collaboration placed an upper bound ${\cal B}(B_s \to \mu^+ \mu^-) <
9.5 \times 10^{-7}$, while the D0 collaboration placed a bound $< 16 \times
10^{-7}$ based on 100 pb$^{-1}$ of data \cite{Nakao:2003}.  The SM prediction
of ${\cal B}(B_s \to \mu^+ \mu^-)$ is more than two orders of magnitude smaller 
than these bounds.

The SM contribution to $B_s \to \mu^+ \mu^-$ is loop-suppressed.  It is
therefore possible for the decay to be dominated by $Z'$ physics.  For an order
of magnitude estimate, one can temporarily ignore the RG running effect at the
$b$-$s$-$Z'$ vertex.  The decay width of $B_s \to \mu^+ \mu^-$ is given by
\bea
{\cal B}(B_s \to \mu^+ \mu^-)
&=& \tau(B_s) \frac{G_F^2}{4\pi} f_{B_s}^2 m_{\mu}^2 m_{B_s}
    \sqrt{1-\frac{4 m_{\mu}^2}{m_{B_s}^2}} |V_{tb}^* V_{ts}|^2 \nn \\
&& \quad \times
    \left\{
    \left| \frac{\alpha}{2\pi \sin^2\theta_W} Y \left(\frac{m_t^2}{M_W^2}\right)
      + 2 \left(\frac{g' M_Z}{g_Y M_{Z'}}\right)^2
        \frac{B^L_{bs} B^L_{\mu\mu}}{V_{tb}^* V_{ts}}
    \right|^2
    + \left| 2 \left(\frac{g' M_Z}{g_Y M_{Z'}}\right)^2
        \frac{B^L_{bs} B^R_{\mu\mu}}{V_{tb}^* V_{ts}} \right|^2
      \right\}~,
\eea
where $B^{L,R}_{\mu\mu}$ is the effective $\mu$-$\mu$-$Z'$ coupling at the weak
scale.  One can find the definition of $Y (m_t^2 / M_W^2)$ in the SM part in
Ref.\ \cite{Buchalla:1993bv}; its value is about $1.05$ here.  No mixing
between $Z$ and $Z'$ is assumed here.  Using the central value of the averaged
$B_s$ lifetime $\tau_{B_s} = 1.439$ ps \cite{LEPBOSC} and $f_{B_s} = 232$ MeV,
we obtain a SM branching ratio of $\simeq 4.1 \times 10^{-9}$.  The current CDF
Run II upper bound thus gives a constraint that
\be
\left\{
    \left| 0.0028
      + \left(\frac{g' M_Z}{g_Y M_{Z'}}\right)^2
        \frac{B^L_{bs} B^L_{\mu\mu}}{V_{tb}^* V_{ts}}
    \right|^2
    + \left| \left(\frac{g' M_Z}{g_Y M_{Z'}}\right)^2
        \frac{B^L_{bs} B^R_{\mu\mu}}{V_{tb}^* V_{ts}} \right|^2
      \right\}^{1/2}
\alt 0.043 ~.
\ee

The weak phase $\phi_L$ associated with $B^L_{sb}$ can in principle be
extracted from the time-dependent $CP$ asymmetry measurement in a fashion
similar to the cases studied in the previous sections, but the relevant
experimental information is not presently available.

We note that the above bound has no direct relation with the couplings relevant
to $b \to s {\bar s} s$ transitions.  The lepton couplings to \zpr\ can be much
smaller than the quark couplings, as is true in some (quasi)leptophobic models.
We are currently investigating the \zpr\ effects on $b \to s \ell^+ \ell^-$
decays, which have been measured to good precision recently and should provide
a tighter bound \cite{BCLL,Buchalla:2000sk}.

%%%%%%%%%%%%%%%%%%%%%%%%%%%%%%%%%%%%%%%%%%%%%%%%%%%%%%%%%%%%%%%%%%%%%%%%%%%%%%%%
\section{Conclusions \label{sec:conclusions}}
%%%%%%%%%%%%%%%%%%%%%%%%%%%%%%%%%%%%%%%%%%%%%%%%%%%%%%%%%%%%%%%%%%%%%%%%%%%%%%%%

In this paper we have considered models with an extra $Z'$ in the mass range of
a few hundred GeV to around 1 TeV.  With a family nonuniversal structure in the
$Z'$ couplings, flavor changing neutral currents are induced via the fermion
mixing, therefore producing interesting effects.  Currently, constraints on the
$Z'$ coupling between the second and third generations are not restrictive.
With non-diagonal left-handed and diagonal right-handed $Z'$ couplings in the
down-type quarks, we studied the impact of such $Z'$ models on rare $B$ meson
decay processes that are sensitive to new physics.

In the present analysis, we have assumed that the left- and right-chiral
couplings $B^L_{ss}$ and $B^R_{ss}$ have the same sign, rendering constructive
interference in the $Z'$ contributions.  We do not include the right-handed
flavor changing couplings, which will give rise to new operators not existent
in the SM.  Involving these or choosing different values for the effective
number of colors $N_C^{\rm eff}$, for which the branching ratios change
sensitively, would change the results.

We have found that with the inclusion of the $Z'$ contributions, $S_{\phi K_S}$
can be appreciably different from the SM prediction, while the branching ratio
of $B^0 \to \phi K^0$ and $A_{\phi K_S}$ are still within the experimental
ranges.  We find that a sizeable weak phase associated with the $B^L_{sb}$
coupling is favored in the ranges of $-80^{\circ} \sim -30^{\circ}$, depending
upon the $\xi^{LL}$ and $\xi^{LR}$ parameter choices.

We have also studied the influence of the new $Z'$ on the $B^0 \to \eta' K^0$
decay.  The $A_{\eta' K_S}$ data do not restrict the choice of $\phi_L$.  The
$S_{\eta' K_S}$ constraint from the data can be satisfied if large couplings
are taken.  Though the discrepancy between the observed branching ratio and the
SM prediction can be explained with this $Z'$ effect, we cannot explain both
branching ratio and CP asymmetries constraints with a common weak angle.
Combining with the constraints from the $B^0 \to \phi K^0$ decays, $S_{\eta'
  K_S}$ and $A_{\eta' K_S}$, we find that a value of $\phi_L$ around
$-60^{\circ} \sim -40^{\circ}$ is favored.

We have observed that the $CP$ asymmetries of the $\phi K_S$ mode are more
sensitive to the \zpr\ effects than the $\eta' K_S$ decay.  This is because of
a cancellation between different parts ($s {\bar s}$ versus $u {\bar u}$ and $d
{\bar d}$) in the $\eta'$ wavefunction.

Finally, we have investigated $B_s \to \mu^+ \mu^-$ decay, now being searched
for at Fermilab, in the same $Z'$ model.  This process can be dominated by the
$Z'$ contribution and the branching ratio can reach the expected sensitivity in
Run II.

C.-W.~C. would like to thank A.~Kagan, D.~Morrissey, J.~Rosner, C.~Wagner, and
L.~Wolfenstein for useful discussions and comments.  This work was supported in
part by the United States Department of Energy under Grants No. EY-76-02-3071
and No. DE-FG02-95ER40896, and in part by the Wisconsin Alumini Research
Foundation.

\end{document}